IAC-22-72306

# Developing a data fusion concept for radar and optical ground based SST station.

**Bruno Coelho[a]\*, Domingos Barbosa[a], Miguel Bergano[a], João Pandeirada[a,b], Paulo Marques[c], Alexandre C. M. Correia[d,e], José Matias de Freitas[f]**

[a] *Instituto de Telecomunicações, Universidade de Aveiro, Campus Universitário de Santiago, 3810-193 Aveiro, Portugal,* brunodfcoelho@av.it.pt, dbarbosa@av.it.pt, jbergano@av.it.pt, joao.pandeirada@av.it.pt
[b] *Departmento de Engenharia Electrónica e de Computadores, Instituto Superior Técnico, Lisboa, Portugal*
[c] *Instituto de Telecomunicações / DEETC-ISEL, IPL, Conselheiro Emídio Navarro 1, 1959-007 Lisboa, Portugal*
[d] *CFisUC, Departamento de Física, Universidade de Coimbra, 3004-516 Coimbra, Portugal*
[e] *ASD, IMCCE-CNRS UMR8028, Observatoire de Paris, PSL Université, 77 Av. Denfert-Rochereau, 75014 Paris, France*
[f] *DGRDN, Ministério da Defesa Nacional, Av. Ilha da Madeira 1, 1400-204 Lisboa, Portugal*
\* Corresponding Author

**Abstract**

As part of the Portuguese Space Surveillance and Tracking (SST) program, a tracking radar and a double Wide Field of View Telescope system (4.3° x 2.3°) are being installed at the Pampilhosa da Serra Space Observatory (PASO) in the centre of continental Portugal, complementing an already installed deployable optical sensor for MEO and GEO surveillance. The tracking radar will track space debris in Low Earth Orbit (LEO) up to 1000 km and at the same time the telescope will also have LEO tracking capabilities.

This article intends to discuss possible ways to take advantage of having these two sensors at the same location. Using both types of sensors takes advantage of the radar measurements which give precise radial velocity and distance to the objects, while the telescope gives better sky coordinates measurements. With the installation of radar and optical sensors, PASO can extend observation time of space debris and correlate information from optical and radar provenances in real time. During twilight periods both sensors can be used simultaneously to rapidly compute new TLEs for LEO objects, eliminating the time delays involved in data exchange between sites in a large SST network. This concept will not replace the need for a SST network with sensors in multiple locations around the globe, but will provide a more complete set of measurements from a given object passage, and therefore increase the added value for initial orbit determination, or monitoring of reentry campaigns of a given location. PASO will contribute to the development of new solutions to better characterize the objects improving the overall SST capabilities and constitute a perfect site for the development and testing of new radar and optical data fusion algorithms and techniques for space debris monitoring.

**Acronyms/Abbreviations**

Space Surveillance and Tracking (SST), Low Earth Orbit (LEO), Medium Earth Orbit (MEO), Geostationary Orbit (GEO), Pampilhosa da Serra Space Observatory (PASO), rAdio TeLescope pAmpilhosa Serra (ATLAS), Field of View (FoV), Portuguese Ministry of Defense (MoD), Radar Cross Section (RCS), Instituto de Telecomunicações (IT), Two-Line Element (TLE)

**1. Introduction**

The number of space debris has been increasing yearly with the progress of many country's space sectors and rise in satellite launches. This growing population of space debris increases the potential danger to all space vehicles and in-space infrastructures, from expensive communications satellites, Earth Observation satellite constellations, to the International Space Station, space shuttles and other spacecraft with humans aboard. For this reason, it is important to create a set of preventive measures in order to avoid any damages to space satellites. In particular, the orbit occupancy in LEO and MEO orbits faces big challenges with planned deployment of space mega-constellations requiring constant and extensive monitoring efforts to avoid collisions. Thus, a network of radar and optical sensors do offer a very high value service to deliver good performance in monitoring and tracking debris of many sizes and in a wide orbit range. [1,2,3].

Radars provide crucial information about the distance and radial velocity components, on the other hand telescopes provide high accuracy. Combining the data from the two types of sensors is crucial for better orbit determination. Usually radars and telescopes are at different location, and are operated by different institutions. This may imply time delays or high latency in data exchanges between sites. This fact, makes it difficult to obtain both types of data in remote locations for a given passage of a putative LEO object. This work pinpoints some advantages of having both kinds of sensors in the same location, describing the infrastructure being deployed at PASO.

## 2. PASO

The Pampilhosa da Serra Space Observatory (PASO) is located in the center of the continental Portuguese territory, in the heart of a certified Dark Sky area. As part of the Portuguese Space Surveillance & Tracking (SST) network, PASO hosts the rAdio TeLescope pAmpilhosa Serra (ATLAS) currently in test phase, and a wide field of view telescope system still in development.

*2.1 Radar*

As part of the Portuguese SST project, led by the Portuguese Ministry of Defense (MoD), Instituto de Telecomunicações (IT) is developing ATLAS, a monostatic radar tracking sensor. The system operates at 5.56 GHz and aims to provide range and range-rate measurements of objects in LEO with RCS above 10 cm² at 1000 km. The radar system was recently installed at PASO, on the 9 m Cassegrain antenna (see Fig. 1) with a beamwidth of (0.73 degrees), and is now in field test phase. For more details on ATLAS see Pandeirada et al. (IAC-22-72291) and [4].

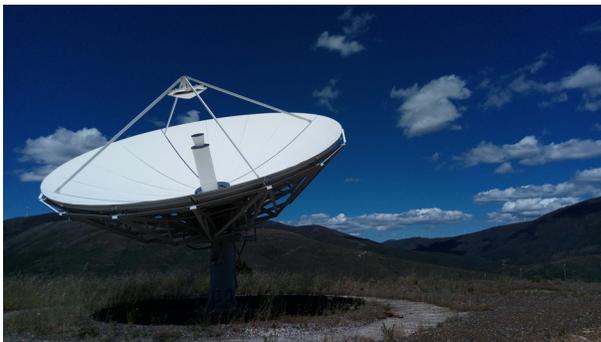

Fig. 1. ATLAS the 5.56 GHz monostatic radar tracking sensor located at PASO.

*2.2 Wide field of view telescope system*

Also recently, was installed at PASO a double wide field of view telescope system with a maximum FoV of 4.3° x 2.3° (see Fig. 2). Its deployment results from a partnership between Portuguese MoD, IT and Universidade de Coimbra through the CfisUC, Departamento de Física. The sensor consists of two small aperture (30cm) telescopes. The system can observe in white light, in BVRI bands, and in the narrow bands H(alpha) and O[III]. The equatorial mount has very fast speeds (max. slewing speed of 40°/s) enabling tracking capability of LEO objects. PASO location has excellent all sky clearance, in a hilltop at 840 m, and surrounded by mountain ranges that allow protection from the light pollution of the coastal regions of the country. The conditions are among the best in continental Portugal for optical use in SST operations, where sky background reaches mag 21 or above and with more than 200 clear nights per year [5].

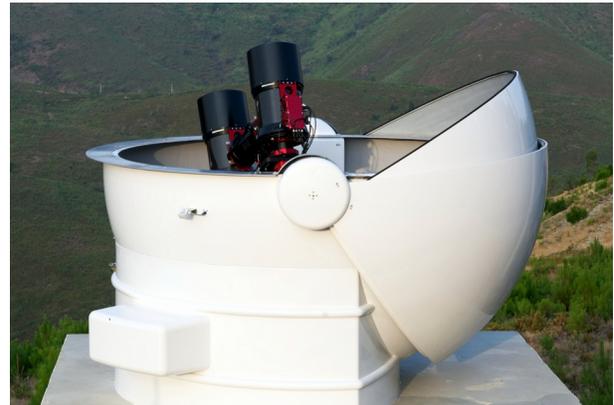

Fig. 2. Double wide field of view telescope system (4.3° x 2.3°) located at PASO.

## 3. Developing a data fusion concept at PASO

Radars can extend the period of SST operations, since they do not suffer weather constrains, however they are not so effective for objects higher than the lower-MEO region. Optical sensors can observe objects in LEO up to GEO regimes, but their operation depends on weather conditions, they can just be used during night time and depending also on the illumination condition of the objects. Radar measurements give precise radial velocity and distance to the objects, on the other hand telescopes give better sky coordinates measurements. The fusion of radar and optical data constitutes then an ideal situation to improve the effective operational status of a SST-network, and allows for better orbit determination [6].

In a large network of sensors, like the EU-SST, there are both types of sensors, but while the telescopes are spread around the world, radar sensors tend to be concentrated in Europe. Since PASO is a ground base equipped with both kind of sensors, it is an excellent opportunitiy to explore the advantages and the new possible methods of integrated sensor operation and data fusion techniques that will include artificial intelligence or machine learning enabled tools.

One very interesting possibility of radar-optical sensor integrated operation was explored in Ladd et al. [7], where an optical sensor was used to track an object during an re-entry, and feed the look angle data to the radar, in real-time. They were able to use the optical system, taking advantage of its larger field of view, to correctly point the radar to it. In that particular case the TLE was not precise enough to get the object inside the radar beamwidth. In such operational model, in PASO the 4.3º x 2.3º FoV of the Double Telescope system can accommodate a larger span of manoeuvres or other changes of the objects trajectories during re-entries than the 0.73° of the radar beamwidth, and that could be the difference between take radar measurements or miss the target due to a not updated TLE. There is also the advantage of simply having both types of measurements of a given object passage.

Having both types of sensor in the same location will be benefitional by allowing the opportunity to acquire both types of data (radar and optical) and by removing the dependence on having different institutions operating the different sensors. This also allows a faster integration of both data types into the data analisys pipelines at space operation centres. This presents an important added value during fast changing scenarios like in large debris re-entries.

At PASO both sensors are so close to each other (see Fig. 3) that they can even use a common time stamping system for the measurements. This situation makes it much easier to coordinate the usage of both sensors to conduct exploratory experiences, and to keep those experiences in a most controlled scenario.

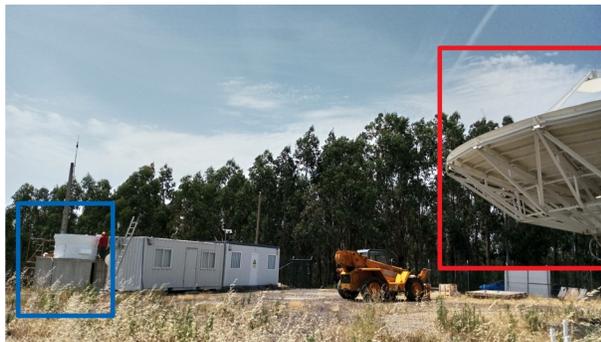

Fig. 3. In this picture taken at PASO it is visible the telescope dome during its installation (in blue), and the ATLAS radar antenna dish (in red). The two sensors are only about 34 m apart.

### 4. Conclusions and Future Work

With the installation of radar and optical sensors, PASO can extend observation time of space debris and correlate information from optical and radar provenances in real time. This will take advantage of the accurate range and range-rate measurements provided by the radar and the more accurate position in sky coordinates of the optical system. The data fusion of both data types is an added value and will contribute to much improve orbit determination. Having both sensors in the same place facilitates coordinated operations, and provide the best conditions to conduct studies on data fusion pipelines and coordinated data acquisition techniques in the future. The usage of the large FoV of the optical system may also increase the productivity of the radar tracking of LEO objects, that usually relies on TLE availability.

### Acknowledgements

The team acknowledges financial support from ENGAGE-SKA Research Infrastructure, ref. POCI-01-0145-FEDER-022217, funded by COMPETE 2020 and FCT, Portugal; from the European Commission H2020 Programme under the grant agreement 2-3SST2018-20. IT team membres acknowledge support from Projecto Lab. Associado UID/EEA/50008/2019.